\documentclass[prl,aps,twocolumn,superscriptaddress]{revtex4}
\usepackage{amsmath}
\usepackage{bm}
\usepackage[normalem]{ulem}
\usepackage{euscript}
\usepackage[pdftex]{graphicx}
\usepackage{xspace}
\usepackage{xfrac}
\usepackage{enumerate}
\usepackage{braket}
\usepackage{float,fancyvrb}
\usepackage[T1]{fontenc}
\usepackage{lmodern}
\graphicspath{{./Figures/}}
\usepackage{wrapfig}
\usepackage{scrextend}
\usepackage{comment}
\usepackage{url}
\usepackage{amssymb}
\usepackage{xcolor} 

\newcommand{\uu}[1]{\ensuremath{\, \mathrm{#1}}} 

\definecolor{ao}{rgb}{0.0, 0.5, 0.0}

\begin{document}
\title{Sensitivity enhancement for a light axion dark matter search with magnetic material}

\author{Alexander V. Gramolin}
\affiliation{Department of Physics, Boston University, Boston, Massachusetts 02215, USA}
\author{Deniz Aybas}
\affiliation{Department of Physics, Boston University, Boston, Massachusetts 02215, USA}
\affiliation{Department of Electrical and Computer Engineering, Boston University, Boston, Massachusetts 02215, USA}
\author{Dorian Johnson}
\affiliation{Department of Physics, Boston University, Boston, Massachusetts 02215, USA}
\author{Janos Adam}
\affiliation{Department of Physics, Boston University, Boston, Massachusetts 02215, USA}
\author{Alexander O. Sushkov}
\email{asu@bu.edu}
\affiliation{Department of Physics, Boston University, Boston, Massachusetts 02215, USA}
\affiliation{Department of Electrical and Computer Engineering, Boston University, Boston, Massachusetts 02215, USA}
\affiliation{Photonics Center, Boston University, Boston, Massachusetts 02215, USA}

\date{\today}

\begin{abstract}
The sensitivity of experimental searches for axion dark matter coupled to photons is typically proportional to the strength of the applied static magnetic field. We demonstrate how a permeable material can be used to enhance the magnitude of this static magnetic field, and therefore improve the sensitivity of such searches in the low frequency lumped-circuit limit. Using gadolinium iron garnet toroids at temperature 4.2~K results in a factor of 4 enhancement compared to an air-core toroidal design. The enhancement is limited by magnetic saturation. Correlation of signals from three such toroids allows efficient rejection of systematics due to electromagnetic interference.
The sensitivity of a centimeter-scale axion dark matter search based on this approach is on the order of $g_{a\gamma\gamma}\approx10^{-9}$~GeV$^{-1}$ after 8 hours of data collection for axion masses near $10^{-10}$~eV. This approach may substantially extend the sensitivity reach of large-volume lumped element axion dark matter searches.
\end{abstract}

\maketitle

Several decades after its theoretical prediction, the axion remains a compelling solution to the strong CP problem, and a well-motivated dark matter candidate~\cite{Abbott1983,Preskill1983,DeMille2017,Irastorza2018a}. Experimental searches for axions rely on one of their interactions with Standard Model particles~\cite{Graham2013,Budker2014,Arvanitaki2014,Abel2017a}. Most experiments to date have focused on the axion-photon interaction, which can convert axions into photons in the presence of strong magnetic field. This effect has been used to place stringent limits on the coupling of axions produced in the laboratory~\cite{Ehret2010} or in the Sun~\cite{Anastassopoulos2017}, approaching the most restrictive astrophysical bounds for a wide range of axion masses~\cite{Payez2015}. 

The technique of resonant conversion of dark matter axions into monochromatic microwave photons inside a high-quality-factor cavity permeated by a strong magnetic field has achieved the level of sensitivity sufficient to search for dark matter axions with masses between 2.66 and $2.81\uu{\mu eV}$, and used to exclude the axion-photon couplings predicted by plausible models for this mass range~\cite{Du2018}. A number of cavity-based axion haloscopes are in development, or already exploring axion frequencies up to $\approx 10$~GHz~\cite{Graham2015a,Brubaker2017}. 
In order to search for lower-mass axions coupled to photons, it is necessary to use lumped-element circuits~\cite{Sikivie2014b,Chaudhuri2015,Kahn2016,Choi2017b,Chaudhuri2018,Ouellet2018}. The experimental concept is based on an effective modification of Maxwell's equations: in presence of a large static magnetic field $\vec{B}$, axion dark matter acts as a source of an oscillating magnetic field, with magnitude proportional to the axion-photon coupling strength $g_{a\gamma\gamma}$, and with oscillation frequency $\omega_a=m_ac^2/\hbar$, where $m_a$ is the axion mass, $c$ is the speed of light, and $\hbar$ is the Planck constant. The coherence time of this oscillating field is $10^6$ periods, set by the kinetic energy of the virialized axion dark matter~\cite{Sikivie1983,Graham2013}.

Our approach is to use toroidally-shaped permeable material to enhance the magnitude of the static magnetic field $\vec{B}$. The inhomogeneous magnetic Maxwell's equation with axion-photon coupling takes the form 
\begin{align}
\vec{\nabla}\times\vec{B} -\frac{1}{c^2}\frac{\partial\vec{E}}{\partial t}= \mu_0 \vec{j}_{\text{el}} + \frac{g_{a\gamma\gamma}}{c}\frac{\partial a}{\partial t}\vec{B},
\label{eq:211}
\end{align}
where $\vec{j}_{\text{el}}$ is the electric current density, and $a=a_0\sin\omega_a t$ is the axion field~\cite{Sikivie1983,Wilczek1987}. We use SI units for electromagnetic fields and natural units for the axion field, so that $g_{a\gamma\gamma}a_0$ is unitless, and the axion field amplitude is given by $m_a^2a_0^2/2=\rho_{\text{DM}}=3.6\times10^{-42}\uu{GeV^4}$~\cite{PDG}. We have neglected the spatial gradient $\vec{\nabla}a$, which is suppressed in the lumped-circuit case. In the presence of a magnetizable medium and macroscopic free electric current density $\vec{J}_f$, this equation takes the form
\begin{align}
\vec{\nabla}\times\vec{H} = \vec{J}_f + \frac{g_{a\gamma\gamma}}{\mu_0c}\frac{\partial a}{\partial t}\vec{B},
\label{eq:211}
\end{align}
where the auxiliary field $\vec{H} = \vec{B}/\mu_0-\vec{M}$ and we assumed vanishing macroscopic charge~\cite{som}. The boundary conditions for fields $\vec{B}$ and $\vec{H}$ at interfaces between different media are unchanged from the usual situation in electromagnetism~\cite{Jackson}.
The enhancement arises from the fact that the magnetic field inside the permeable material includes the material magnetization, in addition to the field created by the free current: $\vec{B}=\mu_0(\vec{H}+\vec{M})$. For a linear magnetic material with permeability $\mu$, $B_0=\mu\mu_0H_0$ and the enhancement is given by $\mu$.

\begin{figure}[h!]
	\centering
	\includegraphics[width=0.9\columnwidth]{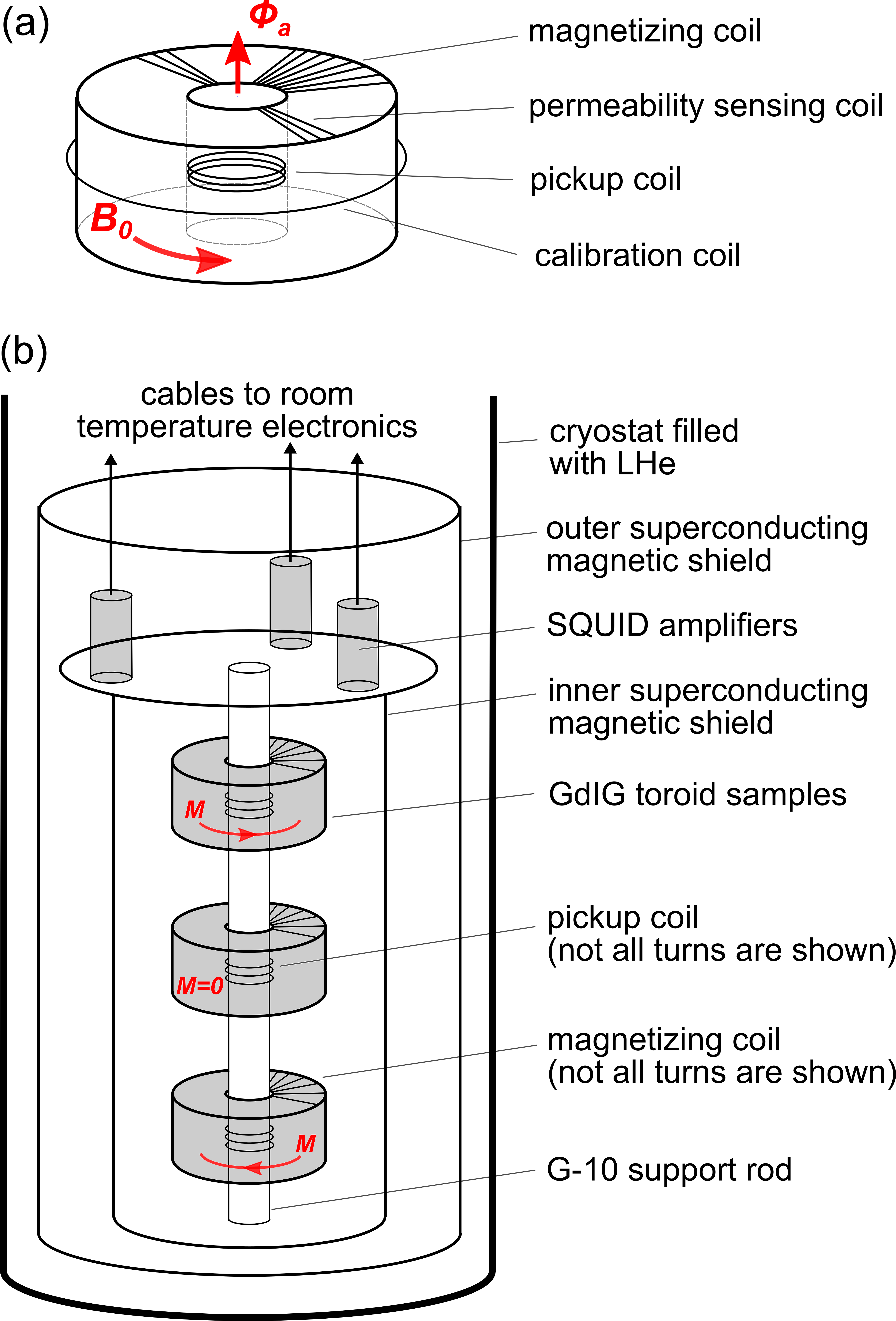}
	\caption{(a) The coils wrapped around the GdIG toroid: a toroidal magnetizing coil created the azimuthal magnetic field $B_0$, a permeability sensing coil was used to monitor the sample magnetization, a pickup coil was sensitive to axion-induced magnetic flux $\Phi_a$, and a calibration coil was used to calibrate the coupling of the pickup coil to the SQUID. Not all coil turns are shown.
	(b) Experimental schematic showing the main components of the apparatus: three GdIG toroids with pickup coils connected to SQUID amplifiers, mounted inside superconducting magnetic shields, immersed in liquid helium at temperature 4.2~K.}
	\label{fig:1}
\end{figure}

The toroidal samples used in our measurements were made of gadolinium iron garnet ferrite (GdIG, chemical formula Gd$_3$Fe$_5$O$_{12}$). This material was chosen for its high permeability at liquid helium temperature, its good insulating properties that ensured no screening of radiofrequency magnetic fields due to skin effect, and for its well-studied magnetic noise properties~\cite{Eckel2009,Sushkov2009}. The toroids had the following  dimensions: inner radius 1.95~cm, outer radius 3.6~cm, height 2.0~cm~\cite{som}.
A magnetizing coil was wound around each of the ferrite toroids, fig.~\ref{fig:1}(a).
Injecting a current into this coil created an azimuthal magnetic field inside the sample, with magnitude $B_0=\mu_0(H_0+M)$, where $\mu_0H_0$ is the magnetic field that would be created by this current in the absence of permeable material and $M$ is the material's magnetization. In the presence of azimuthal magnetic field $B_0$, the axion dark matter field acts as a source of an oscillating effective azimuthal current density 
\begin{equation}
J_{\text{eff}} = \frac{\omega_a}{\mu_0 c} \, g_{a\gamma\gamma} a_0B_0 \cos{(\omega_a t)},
\label{eq:277}
\end{equation}
which can be used to calculate the resulting axial magnetic field and the magnetic flux $\Phi_a$ through the center of the toroid~\cite{som}.

The experimental apparatus contained three ferrite toroids inside an enclosure formed by nested coaxial cylinders, immersed in liquid helium, fig.~\ref{fig:1}(b). During data collection, each of the three toroids had a different value of the azimuthal magnetic field $B_0$, which allowed us to identify and reject a number of systematic signals caused by electromagnetic interference in the apparatus.
Lead foil, affixed to the inner surfaces of the cylinders and their caps, formed a double-layer superconducting magnetic shield. The three GdIG toroids were mounted inside the inner shield. The top and bottom toroids had magnetizing coils, wound using $N_m\approx 1100 $ turns of NbTi wire in two counter-wound layers. These coils were connected to persistent switch heaters inside the outer magnetic shield to enable persistent current operation. A 60-turn permeability sensing coil was wound symmetrically through the top and bottom toroids, as shown in fig.~\ref{fig:1}(a). These coils were used to monitor the magnetization of the corresponding toroid during data collection~\cite{som}.
Three pickup coils were wound on the G-10 support rod, each with $N_p=8$ turns of NbTi wire, and positioned at the inside circumference of each toroid. A twisted pair connected each pickup coil to a Magnicon superconducting quantum interference device (SQUID) amplifier, mounted inside the outer magnetic shield.  

Several calibration measurements had to be performed before collecting data in the configuration sensitive to axion dark matter. 
The sensitivity of the SQUID detectors was calibrated by injecting a known current into the calibration coil, wound around the outer circumference of the GdIG toroid, and recording the SQUID voltage response. Having measured the mutual inductance between the calibration and the SQUID pickup coils at room temperature, the flux-to-current conversion was extracted for each SQUID. This calibration was used to convert the recorded SQUID voltage into the magnetic flux at the pickup coil. Conversion to axion-photon coupling was performed by integrating the effective current density in eq.~(\ref{eq:277}) over the volume of the toroid~\cite{som}.

\begin{figure}[h!]
	\centering
	\includegraphics[width=\columnwidth]{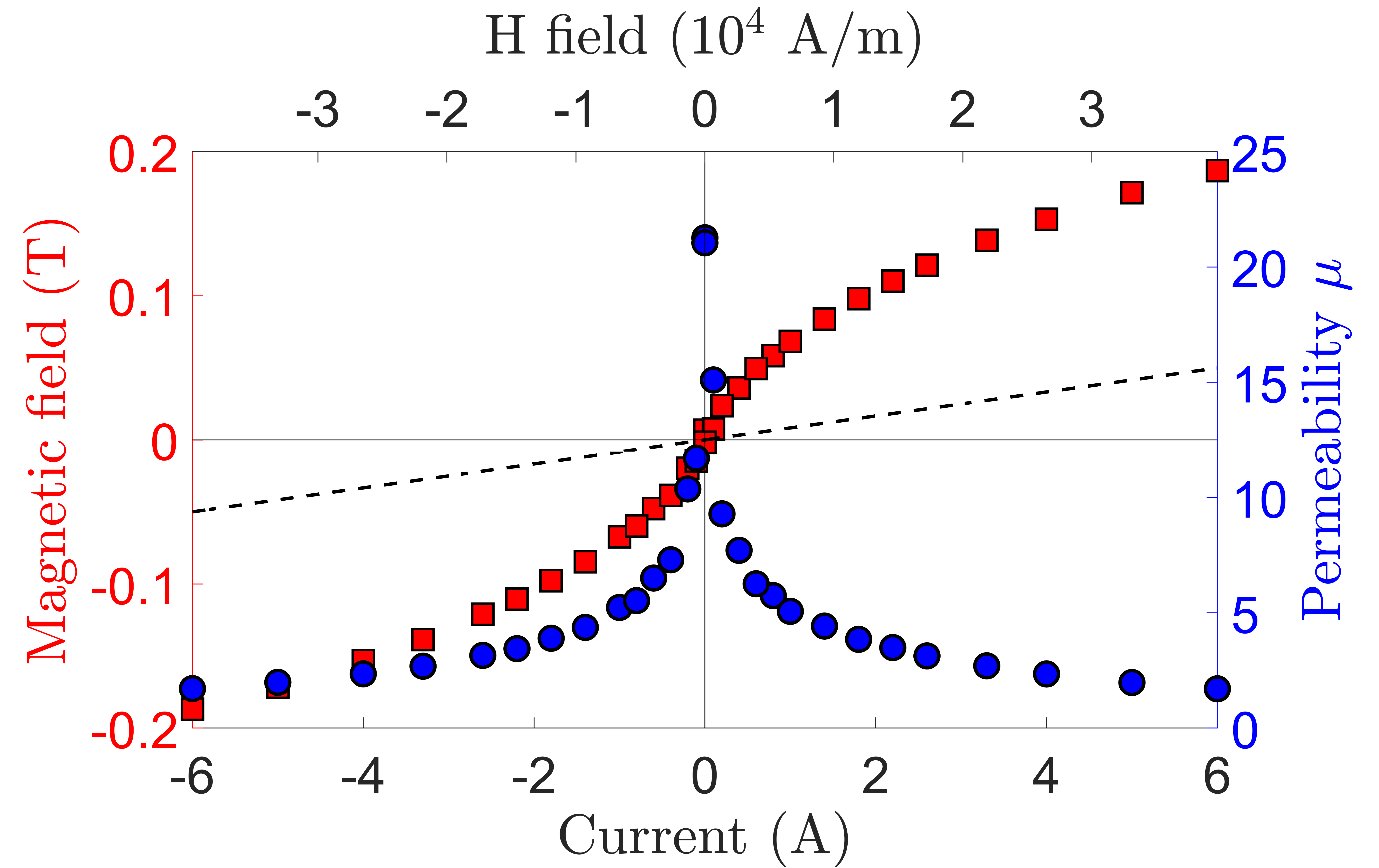}
	\caption{Scaling of GdIG toroid permeability and internal magnetic field with magnetizing coil current $I_m$, at temperature 4.2~K. Blue circles (y-scale on the right) show permeability $\mu=(1/\mu_0)dB/dH$ calculated from measurements of inductance of the magnetizing coil. Red squares (y-scale of the left) show the internal magnetic field calculated using eq.~(\ref{eq:311}). The dashed line shows the magnetic field that would be created inside an air-core toroid.}
	\label{fig:2}
\end{figure}
The azimuthal magnetic field $B_0$ inside the GdIG material was monitored by measuring the inductance of the permeability sensing coil. In order to calibrate this measurement, the inductances of the magnetizing and the permeability sensing coils were simultaneously measured for various values of the constant current through the magnetizing coil~\cite{som}. For each current value, the GdIG permeability $\mu$ was calculated from the inductance of the toroidal magnetizing coil, creating a direct calibration for the inductance of the permeability sensing coil as a function of $\mu$. The initial permeability of the GdIG toroids was $\approx 25$, decreasing with current applied to the magnetizing coil, as the magnetic material saturated, fig.~\ref{fig:2}. Very little magnetic hysteresis was observed.
Since GdIG is a non-linear magnetic material, the value of the azimuthal magnetic field $B_0$ inside the toroid had to be calculated using
\begin{align}
B_0(H_0) = \mu_0\int_0^{H_0}\mu(H)dH.
\label{eq:311}
\end{align} 
After this calibration, a measurement of the inductance of the permeability sensing coil for each toroid gave the value of $B_0$ even when its magnetizing coil was operating in the persistent mode. 
One way to quantify the $B_0$ enhancement is to note that at the injected current of 5~A the magnetic field inside an air-core toroid would be 0.04~T, whereas the magnetic field inside the GdIG toroid was 0.17~T, which is a factor of 4 larger. 

\begin{figure}[h!]
	\centering
	\includegraphics[width=\columnwidth]{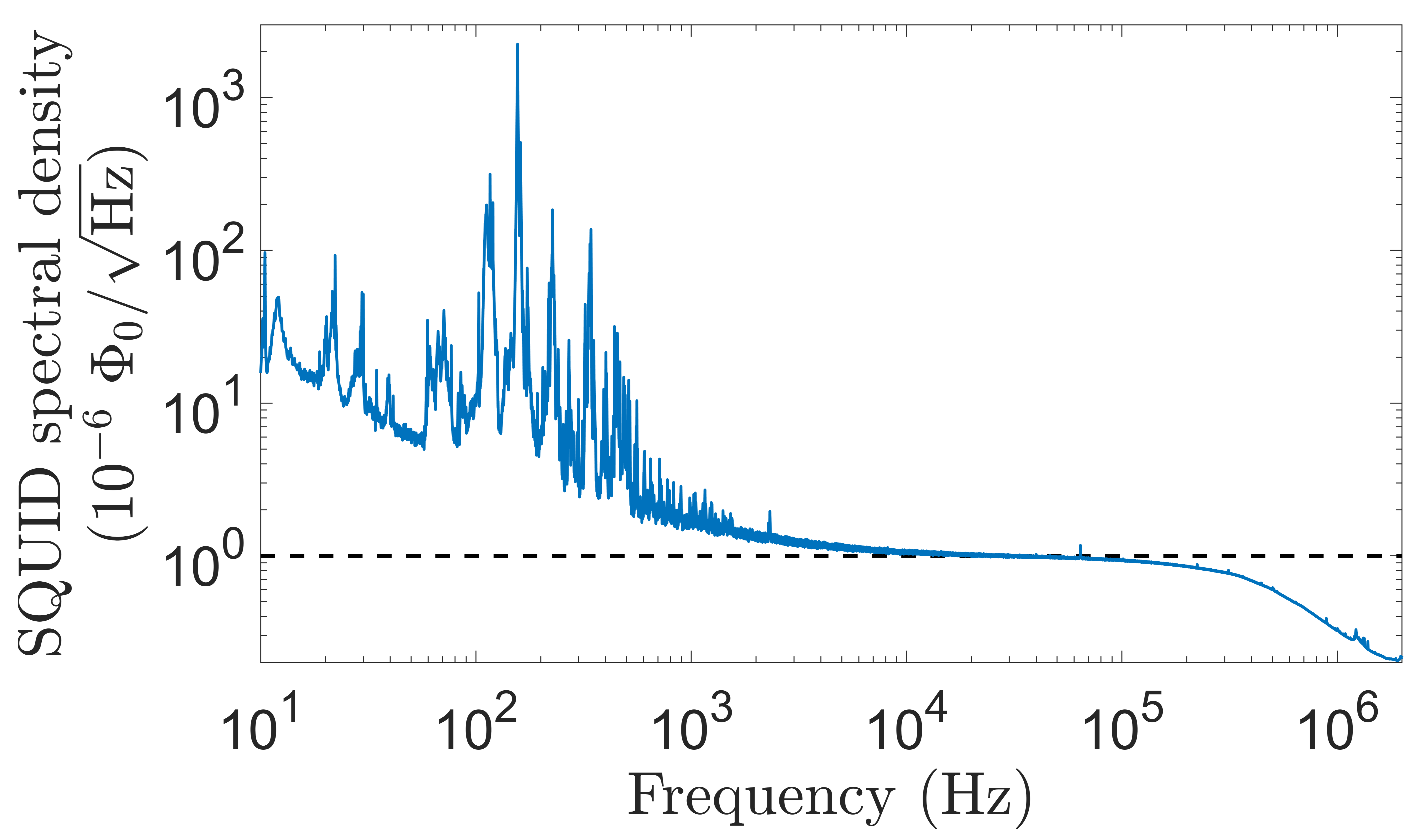}
	\caption{The magnetic flux spectrum recorded by the SQUID amplifier coupled to the bottom GdIG toroid. The dashed line shows the nominal flux noise of the SQUID at $10^{-6}\uu{\Phi_0/\sqrt{Hz}}$, where $\Phi_0$ is the magnetic flux quantum. Noise at frequencies below 1~kHz was correlated with cryostat vibrations. The roll-off at frequencies on the order of MHz is due to the SQUID bandwidth, which was $\approx 400$~kHz for this measurement. }
	\label{fig:3}
\end{figure}

According to the fluctuation-dissipation theorem, the presence of permeable material with complex permeability gives rise to thermal magnetization noise, with a spectrum peaked at low frequencies~\cite{Eckel2009}. Our experimental geometry was designed so that this magnetization noise does not couple to the SQUID pickup coils, which are not sensitive to fluctuations in the toroid magnetization~\cite{Sushkov2009}. Some small coupling is inevitable due to fabrication imperfections, but the magnitude and the $1/f$ power spectral density of the magnetization noise ensure that the dominant noise in our experiment is the intrinsic SQUID detector noise at frequencies above 1~kHz, fig.~\ref{fig:3}. The noise at frequencies below 1~kHz was caused by mechanical vibrations of the cryostat.

In order to evaluate the sensitivity of this experimental approach, we collected data in the setup configuration sensitive to axion dark matter: the middle toroid was unmagnetized, a 1~A current was injected into the magnetizing coil of the top toroid, and a 5~A current was injected into the magnetizing coil of the bottom toroid, such that the magnetization was in the opposite direction. These persistent currents were locked into the coils with corresponding persistent switches. Recorded data spanned a single overnight run with a total integration time $T=28793\uu{s}\approx 8\uu{hours}$. Each of the toroid magnetizations was verified using the corresponding permeability sensing coil after data collection. Signals from the three SQUID detectors were simultaneously digitized at $7.8125\uu{MHz}$ sampling frequency and stored on a hard drive for later analysis.

After performing Fourier transformation and extracting the power spectral density, the spectra were averaged and the data points were grouped into bins, with each bin width set equal to central bin frequency divided by $10^6$ (number of coherent oscillations of the axion field). The axion signal sensitivity at 95\% confidence level at each bin frequency was set to $2\sigma$, where $\sigma$ was the standard deviation of points within that bin. This is a straightforward data analysis approach that gives an estimate for the experimental sensitivity for a basic axion halo model~\cite{Foster2018}.

Electromagnetic interference was observed as peaks in the Fourier spectra of the SQUID detectors at a number of frequencies. Identification and rejection of these systematics was made possible by the simultaneous sampling of signals from the three toroids. If a peak was observed at the same frequency for all three toroids, it was identified as interference, since axion-induced signal would not appear for the middle toroid, which had not been magnetized. The corresponding data points were then removed from the spectrum. Furthermore, an axion peak would appear at different amplitudes in the top and bottom toroids, since they were magnetized with 1~A and 5~A currents, respectively. Therefore,
when a peak was observed for both the top and bottom toroids, its amplitude was scaled by the respective magnetization and the two signals were compared; the axion coupling limit was set to the lower of the two channels~\cite{som}.

\begin{figure}[h!]
	\centering
	\includegraphics[width=\columnwidth]{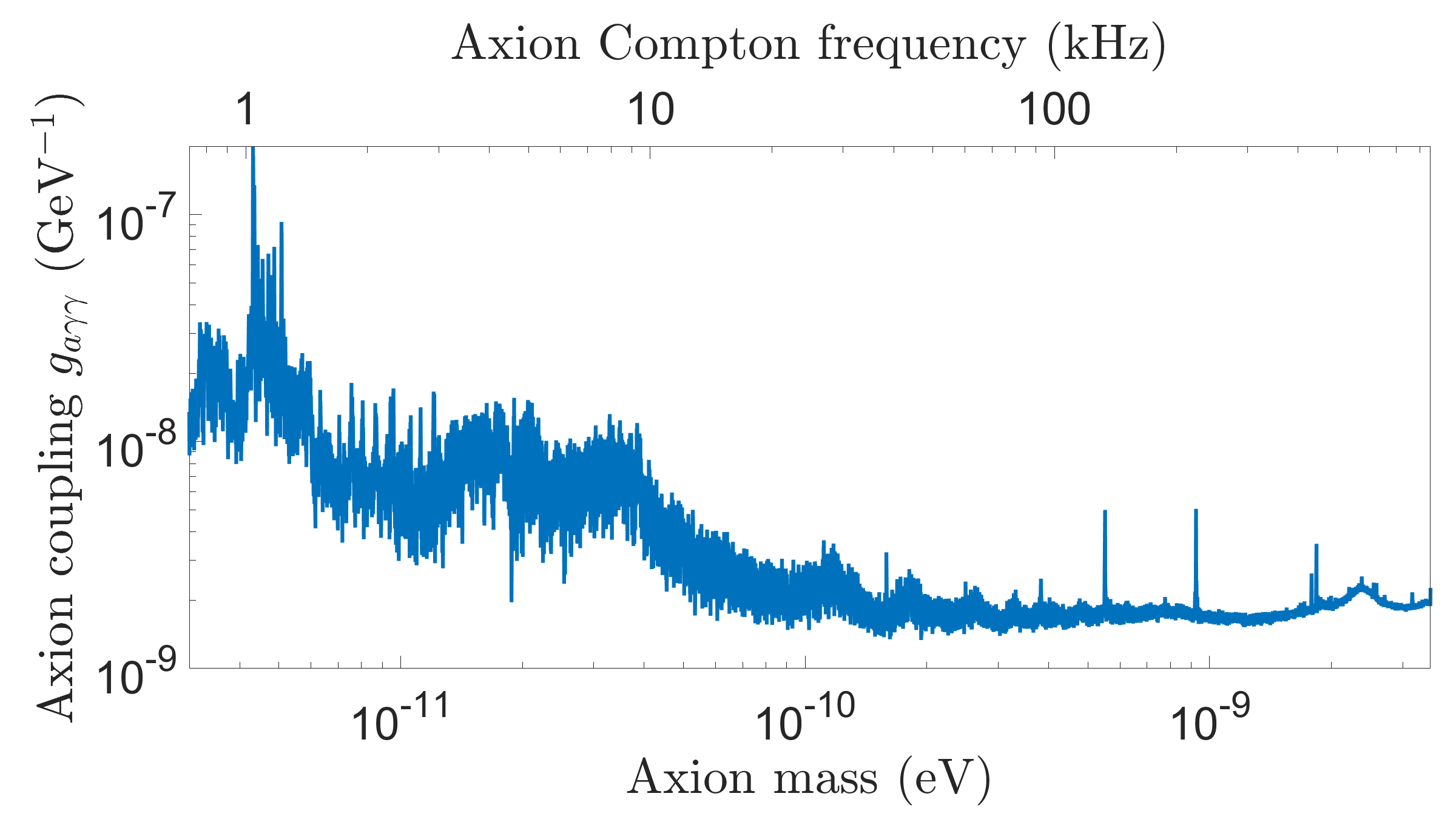}
	\caption{The sensitivity of our centimeter-scale experiment to axion-photon coupling over three orders of magnitude in axion mass between $3\times10^{-12}\uu{eV}$ and $3\times10^{-9}\uu{eV}$. }
	\label{fig:4}
\end{figure}
After 8 hours of data collection at 4.2~K, the sensitivity to axion-photon coupling of a centimeter-scale experimental search for axion dark matter using GdIG was on the order of $10^{-9}\uu{GeV^{-1}}$ in a wide range of axion masses near $10^{-10}\uu{eV}$, fig.~\ref{fig:4}. Over most of the mass range the sensitivity was limited by the SQUID amplifier noise. Cryostat vibrations degrade sensitivity below 20~kHz. The four peaks visible at frequencies 133.27~kHz, 223.72~kHz, 431.91~kHz, and 444.16~kHz could not be rejected with our systematic identification procedure~\cite{som}.
The spectral linewidths of these peaks were larger than what was expected for an axion signal, therefore they were identified as systematics, and sensitivity was degraded near those frequencies. 
The best laboratory limit on axion-photon coupling for this mass range is $0.66\times 10^{-10}\uu{GeV^{-1}}$~\cite{Anastassopoulos2017}.

Using toroidal samples made of permeable material enhanced experimental sensitivity compared to air-core toroidal design by a factor of $\approx 4$, limited by magnetic saturation of GdIG. Adopting other insulating magnetic materials, such as iron powder toroids, that have higher permeability and saturate at much higher fields, may deliver enhancement factors up to a factor of 10 larger. 
The sensitivity of experimental searches for axion dark matter scales with the volume of the region containing static magnetic field. In our geometry the physical volume is $\approx 50\uu{cm^3}$. Scaling this up to the benchmark $1\uu{m^3}$ would improve the sensitivity by another factor of $10^4$~\cite{Sikivie2014b,Chaudhuri2015,Kahn2016}.
Our approach lowers the current required to achieve a given static magnetic field and may substantially simplify design and operation of large-volume superconducting magnets, necessary for such axion dark matter searches. The trade-off is the need for a large magnetic sample, however several promising magnetic materials are used in industrial transformers and are therefore cheap and readily available in large volumes. 
Performing the experiment at a small magnetizing current of 5~A enabled our design with three toroidal samples, allowing efficient identification and rejection of systematics due to electromagnetic interference. Coherent correlation of signals from two toroidal samples with equal and opposite magnetizations may enable an even more effective systematic rejection scheme.

The authors acknowledge support from the NSF grant 1806557, the Heising-Simons and the Simons Foundations grant 2015-039, and the Alfred P. Sloan foundation grant FG-2016-6728.

\bibliography{library} 

\begin{thebibliography}{28}
\expandafter\ifx\csname natexlab\endcsname\relax\def\natexlab#1{#1}\fi
\expandafter\ifx\csname bibnamefont\endcsname\relax
  \def\bibnamefont#1{#1}\fi
\expandafter\ifx\csname bibfnamefont\endcsname\relax
  \def\bibfnamefont#1{#1}\fi
\expandafter\ifx\csname citenamefont\endcsname\relax
  \def\citenamefont#1{#1}\fi
\expandafter\ifx\csname url\endcsname\relax
  \def\url#1{\texttt{#1}}\fi
\expandafter\ifx\csname urlprefix\endcsname\relax\def\urlprefix{URL }\fi
\providecommand{\bibinfo}[2]{#2}
\providecommand{\eprint}[2][]{\url{#2}}

\bibitem[{\citenamefont{Abbott and Sikivie}(1983)}]{Abbott1983}
\bibinfo{author}{\bibfnamefont{L.}~\bibnamefont{Abbott}} \bibnamefont{and}
  \bibinfo{author}{\bibfnamefont{P.}~\bibnamefont{Sikivie}},
  \bibinfo{journal}{Physics Letters B} \textbf{\bibinfo{volume}{120}},
  \bibinfo{pages}{133} (\bibinfo{year}{1983}).

\bibitem[{\citenamefont{Preskill et~al.}(1983)\citenamefont{Preskill, Wise, and
  Wilczek}}]{Preskill1983}
\bibinfo{author}{\bibfnamefont{J.}~\bibnamefont{Preskill}},
  \bibinfo{author}{\bibfnamefont{M.~B.} \bibnamefont{Wise}}, \bibnamefont{and}
  \bibinfo{author}{\bibfnamefont{F.}~\bibnamefont{Wilczek}},
  \bibinfo{journal}{Physics Letters B} \textbf{\bibinfo{volume}{120}},
  \bibinfo{pages}{127} (\bibinfo{year}{1983}).

\bibitem[{\citenamefont{DeMille et~al.}(2017)\citenamefont{DeMille, Doyle, and
  Sushkov}}]{DeMille2017}
\bibinfo{author}{\bibfnamefont{D.}~\bibnamefont{DeMille}},
  \bibinfo{author}{\bibfnamefont{J.~M.} \bibnamefont{Doyle}}, \bibnamefont{and}
  \bibinfo{author}{\bibfnamefont{A.~O.} \bibnamefont{Sushkov}},
  \bibinfo{journal}{Science} \textbf{\bibinfo{volume}{357}},
  \bibinfo{pages}{990} (\bibinfo{year}{2017}).

\bibitem[{\citenamefont{Irastorza and Redondo}(2018)}]{Irastorza2018a}
\bibinfo{author}{\bibfnamefont{I.~G.} \bibnamefont{Irastorza}}
  \bibnamefont{and} \bibinfo{author}{\bibfnamefont{J.}~\bibnamefont{Redondo}},
  \bibinfo{journal}{Progress in Particle and Nuclear Physics}
  \textbf{\bibinfo{volume}{102}}, \bibinfo{pages}{89} (\bibinfo{year}{2018}).

\bibitem[{\citenamefont{Graham and Rajendran}(2013)}]{Graham2013}
\bibinfo{author}{\bibfnamefont{P.~W.} \bibnamefont{Graham}} \bibnamefont{and}
  \bibinfo{author}{\bibfnamefont{S.}~\bibnamefont{Rajendran}},
  \bibinfo{journal}{Physical Review D} \textbf{\bibinfo{volume}{88}},
  \bibinfo{pages}{035023} (\bibinfo{year}{2013}).

\bibitem[{\citenamefont{Budker et~al.}(2014)\citenamefont{Budker, Graham,
  Ledbetter, Rajendran, and Sushkov}}]{Budker2014}
\bibinfo{author}{\bibfnamefont{D.}~\bibnamefont{Budker}},
  \bibinfo{author}{\bibfnamefont{P.~W.} \bibnamefont{Graham}},
  \bibinfo{author}{\bibfnamefont{M.}~\bibnamefont{Ledbetter}},
  \bibinfo{author}{\bibfnamefont{S.}~\bibnamefont{Rajendran}},
  \bibnamefont{and} \bibinfo{author}{\bibfnamefont{A.~O.}
  \bibnamefont{Sushkov}}, \bibinfo{journal}{Physical Review X}
  \textbf{\bibinfo{volume}{4}}, \bibinfo{pages}{021030} (\bibinfo{year}{2014}).

\bibitem[{\citenamefont{Arvanitaki and Geraci}(2014)}]{Arvanitaki2014}
\bibinfo{author}{\bibfnamefont{A.}~\bibnamefont{Arvanitaki}} \bibnamefont{and}
  \bibinfo{author}{\bibfnamefont{A.~A.} \bibnamefont{Geraci}},
  \bibinfo{journal}{Physical Review Letters} \textbf{\bibinfo{volume}{113}},
  \bibinfo{pages}{161801} (\bibinfo{year}{2014}).

\bibitem[{\citenamefont{Abel et~al.}(2017)\citenamefont{Abel, Ayres, Ban,
  Bison, Bodek, Bondar, Daum, Fairbairn, Flambaum, Geltenbort
  et~al.}}]{Abel2017a}
\bibinfo{author}{\bibfnamefont{C.}~\bibnamefont{Abel}},
  \bibinfo{author}{\bibfnamefont{N.~J.} \bibnamefont{Ayres}},
  \bibinfo{author}{\bibfnamefont{G.}~\bibnamefont{Ban}},
  \bibinfo{author}{\bibfnamefont{G.}~\bibnamefont{Bison}},
  \bibinfo{author}{\bibfnamefont{K.}~\bibnamefont{Bodek}},
  \bibinfo{author}{\bibfnamefont{V.}~\bibnamefont{Bondar}},
  \bibinfo{author}{\bibfnamefont{M.}~\bibnamefont{Daum}},
  \bibinfo{author}{\bibfnamefont{M.}~\bibnamefont{Fairbairn}},
  \bibinfo{author}{\bibfnamefont{V.~V.} \bibnamefont{Flambaum}},
  \bibinfo{author}{\bibfnamefont{P.}~\bibnamefont{Geltenbort}},
  \bibnamefont{et~al.}, \bibinfo{journal}{Physical Review X}
  \textbf{\bibinfo{volume}{7}}, \bibinfo{pages}{041034} (\bibinfo{year}{2017}).

\bibitem[{\citenamefont{Ehret et~al.}(2010)\citenamefont{Ehret, Frede,
  Ghazaryan, Hildebrandt, Knabbe, Kracht, Lindner, List, Meier, Meyer
  et~al.}}]{Ehret2010}
\bibinfo{author}{\bibfnamefont{K.}~\bibnamefont{Ehret}},
  \bibinfo{author}{\bibfnamefont{M.}~\bibnamefont{Frede}},
  \bibinfo{author}{\bibfnamefont{S.}~\bibnamefont{Ghazaryan}},
  \bibinfo{author}{\bibfnamefont{M.}~\bibnamefont{Hildebrandt}},
  \bibinfo{author}{\bibfnamefont{E.~A.} \bibnamefont{Knabbe}},
  \bibinfo{author}{\bibfnamefont{D.}~\bibnamefont{Kracht}},
  \bibinfo{author}{\bibfnamefont{A.}~\bibnamefont{Lindner}},
  \bibinfo{author}{\bibfnamefont{J.}~\bibnamefont{List}},
  \bibinfo{author}{\bibfnamefont{T.}~\bibnamefont{Meier}},
  \bibinfo{author}{\bibfnamefont{N.}~\bibnamefont{Meyer}},
  \bibnamefont{et~al.}, \bibinfo{journal}{Physics Letters, Section B: Nuclear,
  Elementary Particle and High-Energy Physics} \textbf{\bibinfo{volume}{689}},
  \bibinfo{pages}{149} (\bibinfo{year}{2010}).

\bibitem[{\citenamefont{Anastassopoulos
  et~al.}(2017)\citenamefont{Anastassopoulos, Aune, Barth, Belov,
  Br{\"{a}}uninger, Cantatore, Carmona, Castel, Cetin, Christensen
  et~al.}}]{Anastassopoulos2017}
\bibinfo{author}{\bibfnamefont{V.}~\bibnamefont{Anastassopoulos}},
  \bibinfo{author}{\bibfnamefont{S.}~\bibnamefont{Aune}},
  \bibinfo{author}{\bibfnamefont{K.}~\bibnamefont{Barth}},
  \bibinfo{author}{\bibfnamefont{A.}~\bibnamefont{Belov}},
  \bibinfo{author}{\bibfnamefont{H.}~\bibnamefont{Br{\"{a}}uninger}},
  \bibinfo{author}{\bibfnamefont{G.}~\bibnamefont{Cantatore}},
  \bibinfo{author}{\bibfnamefont{J.~M.} \bibnamefont{Carmona}},
  \bibinfo{author}{\bibfnamefont{J.~F.} \bibnamefont{Castel}},
  \bibinfo{author}{\bibfnamefont{S.~A.} \bibnamefont{Cetin}},
  \bibinfo{author}{\bibfnamefont{F.}~\bibnamefont{Christensen}},
  \bibnamefont{et~al.}, \bibinfo{journal}{Nature Physics}
  \textbf{\bibinfo{volume}{13}}, \bibinfo{pages}{584} (\bibinfo{year}{2017}).

\bibitem[{\citenamefont{Payez et~al.}(2015)\citenamefont{Payez, Evoli, Fischer,
  Giannotti, Mirizzi, and Ringwald}}]{Payez2015}
\bibinfo{author}{\bibfnamefont{A.}~\bibnamefont{Payez}},
  \bibinfo{author}{\bibfnamefont{C.}~\bibnamefont{Evoli}},
  \bibinfo{author}{\bibfnamefont{T.}~\bibnamefont{Fischer}},
  \bibinfo{author}{\bibfnamefont{M.}~\bibnamefont{Giannotti}},
  \bibinfo{author}{\bibfnamefont{A.}~\bibnamefont{Mirizzi}}, \bibnamefont{and}
  \bibinfo{author}{\bibfnamefont{A.}~\bibnamefont{Ringwald}},
  \bibinfo{journal}{Journal of Cosmology and Astroparticle Physics}
  \textbf{\bibinfo{volume}{02}}, \bibinfo{pages}{006} (\bibinfo{year}{2015}).

\bibitem[{\citenamefont{Du et~al.}(2018)\citenamefont{Du, Force, Khatiwada,
  Lentz, Ottens, Rosenberg, Rybka, Carosi, Woollett, Bowring et~al.}}]{Du2018}
\bibinfo{author}{\bibfnamefont{N.}~\bibnamefont{Du}},
  \bibinfo{author}{\bibfnamefont{N.}~\bibnamefont{Force}},
  \bibinfo{author}{\bibfnamefont{R.}~\bibnamefont{Khatiwada}},
  \bibinfo{author}{\bibfnamefont{E.}~\bibnamefont{Lentz}},
  \bibinfo{author}{\bibfnamefont{R.}~\bibnamefont{Ottens}},
  \bibinfo{author}{\bibfnamefont{L.~J.} \bibnamefont{Rosenberg}},
  \bibinfo{author}{\bibfnamefont{G.}~\bibnamefont{Rybka}},
  \bibinfo{author}{\bibfnamefont{G.}~\bibnamefont{Carosi}},
  \bibinfo{author}{\bibfnamefont{N.}~\bibnamefont{Woollett}},
  \bibinfo{author}{\bibfnamefont{D.}~\bibnamefont{Bowring}},
  \bibnamefont{et~al.}, \bibinfo{journal}{Physical Review Letters}
  \textbf{\bibinfo{volume}{120}}, \bibinfo{pages}{151301}
  (\bibinfo{year}{2018}).

\bibitem[{\citenamefont{Graham et~al.}(2015)\citenamefont{Graham, Irastorza,
  Lamoreaux, Lindner, and van Bibber}}]{Graham2015a}
\bibinfo{author}{\bibfnamefont{P.~W.} \bibnamefont{Graham}},
  \bibinfo{author}{\bibfnamefont{I.~G.} \bibnamefont{Irastorza}},
  \bibinfo{author}{\bibfnamefont{S.~K.} \bibnamefont{Lamoreaux}},
  \bibinfo{author}{\bibfnamefont{A.}~\bibnamefont{Lindner}}, \bibnamefont{and}
  \bibinfo{author}{\bibfnamefont{K.~A.} \bibnamefont{van Bibber}},
  \bibinfo{journal}{Annual Review of Nuclear and Particle Science}
  \textbf{\bibinfo{volume}{65}}, \bibinfo{pages}{485} (\bibinfo{year}{2015}).

\bibitem[{\citenamefont{Brubaker et~al.}(2017)\citenamefont{Brubaker, Zhong,
  Gurevich, Cahn, Lamoreaux, Simanovskaia, Root, Lewis, {Al Kenany}, Backes
  et~al.}}]{Brubaker2017}
\bibinfo{author}{\bibfnamefont{B.~M.} \bibnamefont{Brubaker}},
  \bibinfo{author}{\bibfnamefont{L.}~\bibnamefont{Zhong}},
  \bibinfo{author}{\bibfnamefont{Y.~V.} \bibnamefont{Gurevich}},
  \bibinfo{author}{\bibfnamefont{S.~B.} \bibnamefont{Cahn}},
  \bibinfo{author}{\bibfnamefont{S.~K.} \bibnamefont{Lamoreaux}},
  \bibinfo{author}{\bibfnamefont{M.}~\bibnamefont{Simanovskaia}},
  \bibinfo{author}{\bibfnamefont{J.~R.} \bibnamefont{Root}},
  \bibinfo{author}{\bibfnamefont{S.~M.} \bibnamefont{Lewis}},
  \bibinfo{author}{\bibfnamefont{S.}~\bibnamefont{{Al Kenany}}},
  \bibinfo{author}{\bibfnamefont{K.~M.} \bibnamefont{Backes}},
  \bibnamefont{et~al.}, \bibinfo{journal}{Physical Review Letters}
  \textbf{\bibinfo{volume}{118}}, \bibinfo{pages}{061302}
  (\bibinfo{year}{2017}).

\bibitem[{\citenamefont{Sikivie et~al.}(2014)\citenamefont{Sikivie, Sullivan,
  and Tanner}}]{Sikivie2014b}
\bibinfo{author}{\bibfnamefont{P.}~\bibnamefont{Sikivie}},
  \bibinfo{author}{\bibfnamefont{N.}~\bibnamefont{Sullivan}}, \bibnamefont{and}
  \bibinfo{author}{\bibfnamefont{D.~B.} \bibnamefont{Tanner}},
  \bibinfo{journal}{Physical Review Letters} \textbf{\bibinfo{volume}{112}},
  \bibinfo{pages}{131301} (\bibinfo{year}{2014}).

\bibitem[{\citenamefont{Chaudhuri et~al.}(2015)\citenamefont{Chaudhuri, Graham,
  Irwin, Mardon, Rajendran, and Zhao}}]{Chaudhuri2015}
\bibinfo{author}{\bibfnamefont{S.}~\bibnamefont{Chaudhuri}},
  \bibinfo{author}{\bibfnamefont{P.~W.} \bibnamefont{Graham}},
  \bibinfo{author}{\bibfnamefont{K.}~\bibnamefont{Irwin}},
  \bibinfo{author}{\bibfnamefont{J.}~\bibnamefont{Mardon}},
  \bibinfo{author}{\bibfnamefont{S.}~\bibnamefont{Rajendran}},
  \bibnamefont{and} \bibinfo{author}{\bibfnamefont{Y.}~\bibnamefont{Zhao}},
  \bibinfo{journal}{Physical Review D} \textbf{\bibinfo{volume}{92}},
  \bibinfo{pages}{075012} (\bibinfo{year}{2015}).

\bibitem[{\citenamefont{Kahn et~al.}(2016)\citenamefont{Kahn, Safdi, and
  Thaler}}]{Kahn2016}
\bibinfo{author}{\bibfnamefont{Y.}~\bibnamefont{Kahn}},
  \bibinfo{author}{\bibfnamefont{B.~R.} \bibnamefont{Safdi}}, \bibnamefont{and}
  \bibinfo{author}{\bibfnamefont{J.}~\bibnamefont{Thaler}},
  \bibinfo{journal}{Physical Review Letters} \textbf{\bibinfo{volume}{117}},
  \bibinfo{pages}{141801} (\bibinfo{year}{2016}).

\bibitem[{\citenamefont{Choi et~al.}(2017)\citenamefont{Choi, Themann, Lee, Ko,
  and Semertzidis}}]{Choi2017b}
\bibinfo{author}{\bibfnamefont{J.}~\bibnamefont{Choi}},
  \bibinfo{author}{\bibfnamefont{H.}~\bibnamefont{Themann}},
  \bibinfo{author}{\bibfnamefont{M.~J.} \bibnamefont{Lee}},
  \bibinfo{author}{\bibfnamefont{B.~R.} \bibnamefont{Ko}}, \bibnamefont{and}
  \bibinfo{author}{\bibfnamefont{Y.~K.} \bibnamefont{Semertzidis}},
  \bibinfo{journal}{Physical Review D} \textbf{\bibinfo{volume}{96}},
  \bibinfo{pages}{061102} (\bibinfo{year}{2017}).

\bibitem[{\citenamefont{Chaudhuri et~al.}(2018)\citenamefont{Chaudhuri, Irwin,
  Graham, and Mardon}}]{Chaudhuri2018}
\bibinfo{author}{\bibfnamefont{S.}~\bibnamefont{Chaudhuri}},
  \bibinfo{author}{\bibfnamefont{K.}~\bibnamefont{Irwin}},
  \bibinfo{author}{\bibfnamefont{P.~W.} \bibnamefont{Graham}},
  \bibnamefont{and} \bibinfo{author}{\bibfnamefont{J.}~\bibnamefont{Mardon}},
  \bibinfo{journal}{aXiV:1803.01627}  (\bibinfo{year}{2018}).

\bibitem[{\citenamefont{Ouellet et~al.}(2018)\citenamefont{Ouellet, Salemi,
  Foster, Henning, Bogorad, Conrad, Formaggio, Kahn, Minervini, Radovinsky
  et~al.}}]{Ouellet2018}
\bibinfo{author}{\bibfnamefont{J.~L.} \bibnamefont{Ouellet}},
  \bibinfo{author}{\bibfnamefont{C.~P.} \bibnamefont{Salemi}},
  \bibinfo{author}{\bibfnamefont{J.~W.} \bibnamefont{Foster}},
  \bibinfo{author}{\bibfnamefont{R.}~\bibnamefont{Henning}},
  \bibinfo{author}{\bibfnamefont{Z.}~\bibnamefont{Bogorad}},
  \bibinfo{author}{\bibfnamefont{J.~M.} \bibnamefont{Conrad}},
  \bibinfo{author}{\bibfnamefont{J.~A.} \bibnamefont{Formaggio}},
  \bibinfo{author}{\bibfnamefont{Y.}~\bibnamefont{Kahn}},
  \bibinfo{author}{\bibfnamefont{J.}~\bibnamefont{Minervini}},
  \bibinfo{author}{\bibfnamefont{A.}~\bibnamefont{Radovinsky}},
  \bibnamefont{et~al.}, \bibinfo{journal}{arXiV:1810.12257}
  (\bibinfo{year}{2018}).

\bibitem[{\citenamefont{Sikivie}(1983)}]{Sikivie1983}
\bibinfo{author}{\bibfnamefont{P.}~\bibnamefont{Sikivie}},
  \bibinfo{journal}{Physical Review Letters} \textbf{\bibinfo{volume}{51}},
  \bibinfo{pages}{1415} (\bibinfo{year}{1983}).

\bibitem[{\citenamefont{Wilczek}(1987)}]{Wilczek1987}
\bibinfo{author}{\bibfnamefont{F.}~\bibnamefont{Wilczek}},
  \bibinfo{journal}{Physical Review Letters} \textbf{\bibinfo{volume}{58}},
  \bibinfo{pages}{1799} (\bibinfo{year}{1987}).

\bibitem[{\citenamefont{{M. Tanabashi et al. (Particle Data
  Group)}}(2018)}]{PDG}
\bibinfo{author}{\bibnamefont{{M. Tanabashi et al. (Particle Data Group)}}},
  \bibinfo{journal}{Phys. Rev. D} \textbf{\bibinfo{volume}{98}},
  \bibinfo{pages}{030001} (\bibinfo{year}{2018}).

\bibitem[{som()}]{som}
\emph{\bibinfo{title}{{See Supplemental Material}}}.

\bibitem[{\citenamefont{Jackson}(1999)}]{Jackson}
\bibinfo{author}{\bibfnamefont{J.~D.} \bibnamefont{Jackson}},
  \emph{\bibinfo{title}{{Classical electrodynamics}}}
  (\bibinfo{publisher}{Wiley {\&} Sons}, \bibinfo{year}{1999}).

\bibitem[{\citenamefont{Eckel et~al.}(2009)\citenamefont{Eckel, Sushkov, and
  Lamoreaux}}]{Eckel2009}
\bibinfo{author}{\bibfnamefont{S.}~\bibnamefont{Eckel}},
  \bibinfo{author}{\bibfnamefont{A.~O.} \bibnamefont{Sushkov}},
  \bibnamefont{and} \bibinfo{author}{\bibfnamefont{S.~K.}
  \bibnamefont{Lamoreaux}}, \bibinfo{journal}{Physical Review B}
  \textbf{\bibinfo{volume}{79}}, \bibinfo{pages}{014422}
  (\bibinfo{year}{2009}).

\bibitem[{\citenamefont{Sushkov et~al.}(2009)\citenamefont{Sushkov, Eckel, and
  Lamoreaux}}]{Sushkov2009}
\bibinfo{author}{\bibfnamefont{A.~O.} \bibnamefont{Sushkov}},
  \bibinfo{author}{\bibfnamefont{S.}~\bibnamefont{Eckel}}, \bibnamefont{and}
  \bibinfo{author}{\bibfnamefont{S.}~\bibnamefont{Lamoreaux}},
  \bibinfo{journal}{Physical Review A} \textbf{\bibinfo{volume}{79}},
  \bibinfo{pages}{022118} (\bibinfo{year}{2009}).

\bibitem[{\citenamefont{Foster et~al.}(2018)\citenamefont{Foster, Rodd, and
  Safdi}}]{Foster2018}
\bibinfo{author}{\bibfnamefont{J.~W.} \bibnamefont{Foster}},
  \bibinfo{author}{\bibfnamefont{N.~L.} \bibnamefont{Rodd}}, \bibnamefont{and}
  \bibinfo{author}{\bibfnamefont{B.~R.} \bibnamefont{Safdi}},
  \bibinfo{journal}{Physical Review D} \textbf{\bibinfo{volume}{97}},
  \bibinfo{pages}{123006} (\bibinfo{year}{2018}).

\end{thebibliography}

\end{document}